\documentclass[12pt]{article}
\usepackage[top=25mm,bottom=30mm,left=24mm,right=26mm]{geometry}
\usepackage{amsmath}
\usepackage{tikz}
\usepackage{cite}
\usepackage{import}
\usepackage[utf8]{inputenc}
\usepackage{longtable}
\usepackage[singlelinecheck=off]{caption}
\usepackage{graphicx}
\usepackage[titletoc,title]{appendix}

\numberwithin{equation}{section}
\numberwithin{table}{section}
\numberwithin{figure}{section}

\title {The Higgs Boson Mass from Three-loop Effective Potential of Massless Standard Model}
\begin{document}
\author{}
\date{\small{H. I. Alrebdi$^{1}$, H. A. Alhendi$^{1,2}$ and T. Barakat$^{1}$\\$^1$Department of Physics and Astronomy, College of Science, King Saud University, P. O. Box 2455, Riyadh 11451, Saudi Arabia\\$^2$King Abdullaziz City of Science and Technology, P. O. Box 6086, Riyadh 11442, Saudi Arabia}}
\maketitle
\begin{abstract}
The effective potential of massless standard model (SM) is calculated up to three-loop order. The stability of the effective potential and the Higgs boson mass are investigated up to three-loop order. We found that, Higgs boson mass $m_{H}$ of one-loop order is large. The two-loop and three-loop results are not appreciably different from each other. The two-loop and three-loop radiative corrections have led to an improvement of Higgs boson mass and the value of the scalar coupling. For the value $m_{t}=170$ GeV at the energy scale $\mu\approx 5.68\times10^2$GeV, we get $m_{H2-loop}\approx 125.4$ GeV. At the energy scale $\mu\geq28\times10^2$, the scalar coupling $\lambda$ at two-loop becomes negative and leads to metastable vacuum while the three-loop level is stable even at high $\mu$ $\approx 10^{19}$ GeV. For the larger $\mu$-range $(3 \times 10^{3} \text{GeV}\leq \mu \leq 20 \times 10^{3} \text{GeV})$ spontaneous symmetry breaking for one-loop and three-loop occurs at approximately the same scalar coupling values. We get $m_{H1-loop}\approx 126.14$ GeV at $\mu \approx 6.5 \times 10^{3}$ GeV while for three-loop $m_{H3-loop}\approx 126$ GeV at $\mu \approx 20 \times 10^{3}$ GeV.
\end{abstract}
\noindent\ {\fontsize{9}{8}\selectfont keywords: Effective potential, Renormalization group equation, Spontaneous symmetry breaking, Higgs boson (mass)}
\\
\noindent\small{[PACS: 11.10.Gh, 11.15.Ex, 11.30.Qc, 14.80.Bn]}
\appendix

\renewcommand{\thesection}{\Roman{section}}
\def\theequation{\arabic{section}.\arabic{equation}}
\vspace{60px}
\def\thetable{\arabic{section}.\arabic{table}}
\def\thefigure{\arabic{section}.\arabic{figure}}
\begin{tikzpicture}

\draw (0,0) -- (4,0);
\end{tikzpicture}\\
{\fontsize{10}{4}\selectfont E-mail: haifa.alrebdi$@$gmail.com, alhendi$@$ksu.edu.sa, tbarakat$@$ksu.edu.sa}
\newpage
\section{Introduction}
\paragraph{ } The currently accepted standard model of elementary particles and their interactions is based on Higgs mechanism of spontaneous symmetry breaking of the local electroweak gauge symmetry that leaves the renormalizability of the model intact \cite{1p}. In this mechanism masses of the quarks, the charged leptons, the weak gauge bosons and the neutral scalar Higgs of the model are generated from the asymmetry of the vacuum. 
\paragraph{ } A convenient tool to investigate the nature of the vacuum structure of a renormalizable quantum field theory and stability of the theory with spontaneous symmetry breaking at zero and finite temperature is the effective potential, which contains all quantum radiative corrections to the tree classical potential. The formalism of the effective potential as an essential tool for the investigation of the spontaneous symmetry breaking was first considered by Goldstone and et al \cite{2p} and Jona-Lasinio \cite{3p}, (for review of early works see for example ref.\cite{4p}). The physical interpretation of the effective potential as a means to explore the vacuum structure of a quantum field theory has been elaborated by Symanzik \cite{5p}, (for review see for example ref.\cite{6p} and ref.\cite{7p}). The vacuum structure of a quantum field theory involves several quantitative calculable quantities such as symmetry and asymmetry of the vacuum, stability and instability of the vacuum, symmetry breaking patterns, restoration of symmetry, induced asymmetry and vacuum energy. 
\paragraph{ } The effective potential at higher order is used to investigate the vacuum structure of spontaneous symmetry breaking through Coleman-Weinberg mechanism \cite{9p} for massless standard model (SM). Starting with a theory massless in the tree-level approximation is an attractive idea where it avoids the problems associated with the negative mass square term.
\paragraph{ } The standard model contains some of free parameters, one of these parameters is the mass of the Higgs boson, as a tree level Higgs scalar coupling $\lambda$  so the mass of the Higgs boson has a priori no constraints from SM theory, and its determination self-consistently poses a challenging problem. In the present work, an attempt is made to determine the Higgs boson mass self-consistently from the renormalization group equations and the effective potential of the model. Investigation the stability of the vacuum leads to determine the Higgs boson mass from the effective potential for each order by using the renormalization condition. This paper is organized as follows, in Sec. II, we apply the renormalization group equation method with use corresponding renormalization group functions to calculate the effective potential up to three-loop order for the massless SM. In Sec. III, we drive a relation that allows to calculate the Higgs boson mass and we discuss the results and investigate the stability of the vacuum structure under radiative quantum corrections up to three-loop order in different mass scale. In addition, we compare our result with some results of other authors employing different approaches. In sec. IV, we present our conclusion.

\section{Calculations of the Effective Potential from the Renormalization Group Equation}
\paragraph{ } In the mass-independent 'tHooft-Weinberg scheme \cite{8p}  the full effective potential $V(\varphi)$ satisfies renormalization group equation RGE. Thus, in general the RGE for $V(\varphi)$:
\begin{equation}
DV(\varphi)=0
\label{eq:2_1}
\end{equation}
where
\begin{equation}
D=\mu\frac{\partial}{\partial \mu}+\beta_{\lambda}\frac{\partial}{\partial \beta_{\lambda}}+m^{2}\beta_{m^{2}}\frac{\partial}{\partial m^{2}}+\beta_{g}\frac{\partial}{\partial g}+\gamma \varphi\frac{\partial}{\partial\varphi}
\label{eq:2_2}
\end{equation}
\\Here $\beta_\lambda$,$\beta_{m^{2}}$,$\beta_{g}$ and $\gamma$ are the renormalization group functions for scalar, mass and gauge couplings, respectively and $\gamma$ is the anomalous dimension, which usually calculated perturbatively. Let us now rewrite the RGE as a short form by defining the parameters of the theory, $\lambda_{p}\equiv(\lambda,m^{2},g)$ to get:
\begin{equation}
\left[\mu\frac{\partial}{\partial \mu}+\delta_{p}\beta_{p}\frac{\partial}{\partial\lambda_{p}}+\gamma\varphi\frac{\partial}{\partial\varphi}\right]V(\varphi)=0
\label{eq:2_3}
\end{equation}
\\
Where $\delta_{p}$ equals $m^{2}$ for mass coupling and 1 otherwise. Also $g$ represents all the gauge couplings $g_{1}$,$g_{2}$,and $g_{3}$ of U(1), SU(2) and SU(3) respectively, in addition to the top quark coupling $g_{t}$. The renormalization group function RGF $\beta_{p}$, which called the beta function is given by:
\begin{equation}
\beta_{p}=\frac{d\lambda_{p}}{dt}, 
\label{eq:2_4}
\end{equation}
with $t=\ln\frac{\mu}{\mu_{0}},\; \mu_{0}=\mu (0)$
\\
In the loop expansion, the effective potential has the expression\cite{9p}:
\begin{equation}
V_{eff}=V^{(0)}+\sum^{\infty}_{n=1}\hbar^{(n)}V^{(n)},
\label{eq:2_5}
\end{equation}
Likewise, the renormalization group functions have the expansion:
\begin{equation}\begin{aligned}
\beta_{p}=\sum^{\infty}_{n=1}\hbar^{(n)}\beta^{(n)}_{p},\\
\gamma=\sum^{\infty}_{n=1}\hbar^{(n)}\gamma^{(n)},
\end{aligned}
\label{eq:2_6}
\end{equation}
where the $\chi^{(n)}$ is the n-loop contribution to $\chi$. Then following the ref.\cite{10p}, we can write the following n-loop of $\hbar^{n}$-order as:
\begin{equation}
\mu\frac{\partial}{\partial\mu}V^{(n)}+D_{n}V^{(0)}+D_{n-1}V^{(1)}+...+D_{1}V^{(n-1)}=0,
\label{eq:2_7}
\end{equation}
where
\begin{equation}
D_{m}=\delta_{p}\beta^{(m)}_{p}\frac{\partial}{\partial\lambda_{p}}+\gamma^{(m)}\varphi\frac{\partial}{\partial\varphi},\;\;m=1,2,...,n
\label{eq:2_8}
\end{equation}
is the differential operator. Hence, one can find the n-loop contribution to the effective potential by using the recursion formula eq.(\ref{eq:2_7}) in terms of RGFs and the tree level potential.
\paragraph{ }The tree level (zero-loop) effective potential for massless SM is:
\begin{equation*}
V^{(0)}=\frac{\lambda}{3!}(\phi^{+}\phi), \text{putting}\;\;\phi^{+}\phi=\frac{\varphi^{2}}{2}, 
\end{equation*}
one obtains:

\begin{equation}
V^{(0)}=\frac{\lambda}{4!}\varphi^{4}
\label{eq:2_9}
\end{equation}
\\Now, applying eq.(\ref{eq:2_7}) to eq.(\ref{eq:2_9}) using eqs.(\ref{eq:2_6}) and (\ref{eq:2_8}) one gets for one-loop contribution to the effective potential:
\begin{equation}
\mu\frac{\partial}{\partial\mu}V^{(1)}+\frac{A_{1}}{4!}\varphi^{4}=0,
\label{eq:2_10}
\end{equation}
where
\begin{equation}
A_{1}=\beta^{(1)}_{\lambda}+4\lambda\gamma^{(1)}.
\label{eq:2_11}
\end{equation}
Integration of eq.(\ref{eq:2_10}) yields:
\begin{equation}
V^{(1)}=-\frac{A_{1}}{48}\varphi^{4}\ln\frac{\mu^{2}}{\mu^{2}_{0}}+C_{1}(\lambda,g,\varphi,\mu_{0}),
\label{eq:2_12}\end{equation}
\\The two-loop contribution to the effective potential:
\begin{equation}
V^{(2)}=-\frac{A}{48}\varphi^{4}\ln\frac{\mu^{2}}{\mu^{2}_{0}}-\frac{B}{96}\varphi^{4}\ln\frac{\varphi^{2}}{\mu^{2}_{0}}\ln\frac{\mu^{2}}{\mu^{2}_{0}}+\frac{B}{192}\varphi^{4}(\ln\frac{\mu^{2}}{\mu^{2}_{0}})^{2}+C_{2}(\lambda,g,\varphi,\mu_{0}),
\label{eq:2_13}\end{equation}
with
\begin{equation}
A=A_{1}\gamma^{(1)}+A_{2}, \;\;A_{2}=\beta^{(2)}_{\lambda}+4\lambda\gamma^{(2)},
\label{eq:2_14}\end{equation}\\
\begin{equation}
B=[\beta^{(1)}_{\lambda}\frac{\partial\beta^{(1)}_{\lambda}}{\partial\lambda}+8\beta^{(1)}_{\lambda}\gamma^{(1)}+16\lambda(\gamma^{(1)})^{2}+F_{1}],
\label{eq:2_15}\end{equation}\\
\begin{equation}\begin{aligned}
F_{1}=[\beta^{(1)}_{g_{1}}\frac{\partial\beta^{(1)}_{\lambda}}{\partial g_{1}}+4\lambda\beta^{(1)}_{g_{1}}\frac{\partial\gamma^{(1)}}{\partial g_{1}}+\beta^{(1)}_{g_{2}}\frac{\partial\beta^{(1)}_{\lambda}}{\partial g_{2}}+4\lambda\beta^{(1)}_{g_{2}}\frac{\partial\gamma^{(1)}}{\partial g_{2}}\\+\beta^{(1)}_{g_{3}}\frac{\partial\beta^{(1)}_{\lambda}}{\partial g_{3}}+4\lambda\beta^{(1)}_{g_{3}}\frac{\partial\gamma^{(1)}}{\partial g_{3}}+\beta^{(1)}_{g_{t}}\frac{\partial\beta^{(1)}_{\lambda}}{\partial g_{t}}+4\lambda\beta^{(1)}_{g_{t}}\frac{\partial\gamma^{(1)}}{\partial g_{t}}].
\end{aligned}
\label{eq:2_16}\end{equation}\\
For the three-loop contribution to the effective potential, one gets:
\begin{equation}\begin{aligned}
V^{(3)}=\frac{-K_{1}}{48}\varphi^{4}\ln\frac{\mu^{2}}{\mu^{2}_{0}}-\frac{K_{2}}{96}\varphi^{4}(\ln\frac{\mu^{2}}{\mu^{2}_{0}}\ln\frac{\varphi^{2}}{\mu^{2}_{0}})+\frac{K_{2}}{192}\varphi^{4}(\ln\frac{\mu^{2}}{\mu^{2}_{0}})^{2}\\-\frac{1}{192}\frac{K_{3}}{2}[\ln\frac{\mu^{2}}{\mu^{2}_{0}}(\ln\frac{\varphi^{2}}{\mu^{2}_{0}})^{2}-\ln\frac{\varphi^{2}}{\mu^{2}_{0}}(\frac{\mu^{2}}{\mu^{2}_{0}})^{2}]-\frac{K_{3}}{1152}\varphi^{4}(\ln\frac{\mu^{2}}{\mu^{2}_{0}})^{3}+C_{3}(\lambda,g,\varphi,\mu_{0}),
\end{aligned}
\label{eq:2_17}\end{equation}
where
\begin{equation}
K_{1}=[A_{1}\gamma^{(2)}+A\gamma^{(1)}+A_{3}],
\label{eq:2_18}\end{equation}
with
\begin{equation}
A_{3}=\beta^{(3)}_{\lambda}+4\lambda\gamma^{(3)},
\label{eq:2_19}\end{equation}
\\
\begin{equation}\begin{aligned}
K_{2}=[\beta^{(1)}_{\lambda}\frac{\partial A}{\partial\lambda}+\beta^{(2)}_{\lambda}\frac{\partial\beta^{(1)}_{\lambda}}{\partial\lambda}+4\beta^{(2)}_{\lambda}\gamma^{(1)}+4\beta^{(1)}_{\lambda}\gamma^{(2)}\\+\gamma^{(1)}(4A+B)+16\lambda\gamma^{(1)}\gamma^{(2)}+F_{2}+F_{3}],
\end{aligned}\label{eq:2_20}\end{equation}
with
\begin{equation}\begin{aligned}
F_{2}=[\beta^{(2)}_{g_{1}}\frac{\partial\beta^{(1)}_{\lambda}}{\partial g_{1}}+4\lambda\beta^{(2)}_{g_{1}}\frac{\partial\gamma^{(1)}}{\partial g_{1}}+\beta^{(2)}_{g_{2}}\frac{\partial\beta^{(1)}_{\lambda}}{\partial g_{2}}+4\lambda\beta^{(2)}_{g_{2}}\frac{\partial\gamma^{(1)}}{\partial g_{2}}\\+\beta^{(2)}_{g_{3}}\frac{\partial\beta^{(1)}_{\lambda}}{\partial g_{3}}+4\lambda\beta^{(2)}_{g_{3}}\frac{\partial\gamma^{(1)}}{\partial g_{3}}+\beta^{(2)}_{g_{t}}\frac{\partial\beta^{(1)}_{\lambda}}{\partial g_{t}}+4\lambda\beta^{(2)}_{g_{t}}\frac{\partial\gamma^{(1)}}{\partial g_{t}}],
\end{aligned}\label{eq:2_21}\end{equation}\\
\begin{equation}
F_{3}=\beta^{(1)}_{g_{1}}\frac{\partial A}{\partial g_{1}}+\beta^{(1)}_{g_{2}}\frac{\partial A}{\partial g_{2}}+\beta^{(1)}_{g_{3}}\frac{\partial A}{\partial g_{3}}+\beta^{(1)}_{g_{t}}\frac{\partial A}{\partial g_{t}},
\label{eq:2_22}\end{equation}\\
\begin{equation}
K_{3}=[\beta^{(1)}_{\lambda}\frac{\partial B}{\partial \lambda}+4B\gamma^{(1)}+F_{4}],
\label{eq:2_23}\end{equation}
with
\begin{equation}
F_{4}=\beta^{(1)}_{g_{1}}\frac{\partial B}{\partial g_{1}}+\beta^{(1)}_{g_{2}}\frac{\partial B}{\partial g_{2}}+\beta^{(1)}_{g_{3}}\frac{\partial B}{\partial g_{3}}+\beta^{(1)}_{g_{t}}\frac{\partial B}{\partial g_{t}}.
\label{eq:2_24}\end{equation}
\paragraph{ } It is worth noting that integrating renormalization group equation for the effective potential in the RGE method a constant $C$ that depending on $\varphi,m,\lambda,g$ and $\mu_{0}$ as it is clear in eqs.(\ref{eq:2_12}), (\ref{eq:2_13}) and (\ref{eq:2_17}) should be fixed. In general, to determine this constant one needs to specify renormalization conditions on the effective potential. There are several renormalization conditions. For example,
\begin{equation*}
V(\varphi^{2}=\mu^{2})=V^{(0)}(\varphi) \;\;\text{or} \;\;\frac{d^{2}V}{d\varphi^{2}}|_{\varphi=0}=m^{2} \;\;\text{and}\;\; \frac{d^{4}V}{d\varphi^{4}}|_{\varphi=\mu}=\lambda.
\end{equation*}
Here the constants $C_{1},C_{2}$  and $C_{3}$ are determined by using the condition:

\begin{equation}
\frac{d^{4}V}{d\varphi^{4}}|_{\varphi=\mu}=\lambda.
\label{eq:2_25}\end{equation}\\
Therefore, the full effective potential approximated to one-loop, two-loop and three-loop respectively becomes:
\begin{equation}
V_{1-loop\;approx}=V^{(0)}+V^{(1)}=\frac{\lambda}{24}\varphi^{4}+\frac{A_{1}}{48}\varphi^{4}[\ln\frac{\varphi^{2}}{\mu^{2}}-\frac{25}{6}],
\label{eq:2_26}\end{equation}\\
\begin{equation}\begin{aligned}
V_{2-loop\;approx}=V^{(0)}+V^{(1)}+V^{(2)}=\frac{\lambda}{24}\varphi^{4}\\+\frac{(A_{1}+A)}{48}\varphi^{4}[\ln\frac{\varphi^{2}}{\mu^{2}}-\frac{25}{6}]+\frac{B}{192}\varphi^{4}[(\ln\frac{\varphi^{2}}{\mu^{2}})^{2}-\frac{35}{3}],
\end{aligned}\label{eq:2_27}\end{equation}\\
\begin{equation}\begin{aligned}
V_{3-loop\;approx}=V^{(0)}+V^{(1)}+V^{(2)}=\frac{\lambda}{24}\varphi^{4}+\frac{I_{1}}{48}\varphi^{4}[\ln\frac{\varphi^{2}}{\mu^{2}}-\frac{25}{6}]\\+\frac{I_{2}}{192}\varphi^{4}[(\ln\frac{\varphi^{2}}{\mu^{2}})^{2}-\frac{35}{3}]+\frac{I_{3}}{1152}\varphi^{4}[(\ln\frac{\varphi^{2}}{\mu^{2}})^{3}-20],
\end{aligned}\label{eq:2_28}\end{equation}
with
\begin{equation}\begin{aligned}
I_{1}=(A_{1}+A+K_{1}),\;\; I_{2}=(B+K_{2}),\;\; I_{3}=K_{3}.
\end{aligned}\label{eq:2_29}\end{equation}\\
Now by using renormalization group functions in the standard model one can obtain the effective potential of the  massless SM for one-loop eq.(\ref{eq:2_26}), two-loop order eq.(\ref{eq:2_27}) and for  three-loop order eq.(\ref{eq:2_28}), these RGFs are given in the Appendix A for one, two and three level \cite{11p,12p,13p,14p,15p,16p,17p}.

\section{Vacuum Stability and the Higgs Boson Mass from the Effective Potential of Massless Standard Model}
\paragraph{ } In the massless case the scalar-field self-interacting coupling constant $\lambda$, is determined from the minimum of the potential by using the tadpole condition:
\begin{equation}
\frac{dV}{d\varphi}|_{\varphi=v}=0.
\label{eq:3_1}
\end{equation}\\
By applying eq.(\ref{eq:3_1}) to one-loop eq.(\ref{eq:2_26}), two-loop eq.(\ref{eq:2_27}) and three-loop eq.(\ref{eq:2_28}) order effective potential we get:
\begin{equation}
\frac{dV_{1-loop\;approx}}{d\varphi}|_{\varphi=v_{1}}=0\Rightarrow \lambda-\frac{11}{6}A_{1}+\frac{A_{1}}{2}(\ln\frac{v^{2}_{1}}{\mu^{2}})=0,
\label{eq:3_2}
\end{equation}
\begin{equation*}
\frac{dV_{2-loop\;approx}}{d\varphi}|_{\varphi=v_{2}}=0\;\;\;\;\;\;\;\;\;\;\;\;\;\;\;\;\;\;\;\;\;\;\;\;\;\;\;\;\;\;\;\;\;\;\;\;\;\;\;\;\;\;\;\;\;\;\;\;\end{equation*}
\begin{equation}\begin{aligned}
\Rightarrow\lambda-\frac{11}{6}(A_{1}+A)+24(\frac{A_{1}+A}{48}+\frac{B}{192})\ln\frac{v^{2}_{2}}{\mu^{2}}+\frac{24B}{192}[(\ln\frac{v^{2}_{2}}{\mu^{2}})^{2}-\frac{35}{3}]=0,
\end{aligned}\label{eq:3_3}
\end{equation}

\begin{equation*}
\frac{dV_{3-loop\;approx}}{d\varphi}|_{\varphi=v_{3}}=0\;\;\;\;\;\;\;\;\;\;\;\;\;\;\;\;\;\;\;\;\;\;\;\;\;\;\;\;\;\;\;\;\;\;\;\;\;\;\;\;\;\;\;\;\;\;\;\;\end{equation*}
\begin{equation}\begin{aligned}
\Rightarrow\lambda-\frac{11}{6}I_{1}+24(\frac{I_{1}}{48}+\frac{I_{2}}{192})\ln\frac{v^{2}_{3}}{\mu^{2}}+\frac{24I_{2}}{192}[(\ln\frac{v^{2}_{3}}{\mu^{2}})^{2}-\frac{35}{3}]+\frac{36I_{3}}{1152}(\ln\frac{v^{2}_{3}}{\mu^{2}})^{2}+\frac{24 I_{3}}{1152}[(\ln\frac{v^{2}_{3}}{\mu^{2}})^{3}-20]=0.
\end{aligned}\label{eq:3_4}
\end{equation}\\

These equations allows us to obtain the scalar Higgs coupling $\lambda$ for each order if the gauge and Yukawa couplings at the renormalization scale are known, where $v_{i}\; (i=1,2,3)$  is the vacuum expectation value. The vacuum expectation values have the expansions up to one-loop, two-loop and three-loop respectively:
\begin{equation*}
v_{1}=v_{0}+\hbar v^{(1)}, v_{0}=0,
\end{equation*}
\begin{equation*}
v_{2}=\hbar v^{(1)}+\hbar^{2} v^{(2)},
\end{equation*}
\begin{equation*}
v_{3}=\hbar v^{(1)}+\hbar^{2} v^{(2)}+\hbar^{3} v^{(3)}.
\end{equation*}\\
In perturbative theory the order n-contribution should be less than $n-1$-order to make the perturbative series convergent. So, $v^{(2)}$  and $v^{(3)}$  contributions are too small compared to the one-loop contribution value, so we consider the approximation $v_{1},v_{2},v_{3}\approx$ experiment value (246.2 GeV).\\\\
Now the Higgs boson mass is calculated from \cite{9p}:
\begin{equation}
m^{2}_{H}=\frac{d^{2}V}{d\varphi^{2}}|_{\varphi=v},
\label{eq:3_5}
\end{equation}
and satisfies the condition of eq.(\ref{eq:3_1}).\\\\
At one-loop order, using eq.(\ref{eq:2_26}), eq.(\ref{eq:3_2}) and eq.(\ref{eq:3_5}):
\begin{equation}\begin{aligned}
m^{2}_{H1-loop}=\frac{d^{2}V_{1-loop\;approx}}{d\varphi^{2}}|_{\varphi=v_{1}}=\frac{A_{1}}{6}v^{2}_{1},
\end{aligned}\label{eq:3_6}
\end{equation}\\
For two-loop order, using eq.(\ref{eq:2_27}), eq.(\ref{eq:3_3}) and eq.(\ref{eq:3_5}):
\begin{equation}\begin{aligned}
m^{2}_{H2-loop}=\frac{d^{2}V_{2-loop\;approx}}{d\varphi^{2}}|_{\varphi=v_{2}}=(\frac{A_{1}+A}{6}+\frac{B}{24}+\frac{B}{12}\ln\frac{v^{2}_{2}}{\mu^{2}})v^{2}_{2},
\end{aligned}\label{eq:3_7}
\end{equation}\\
and finally at three-loop order, using eq.(\ref{eq:2_28}), eq.(\ref{eq:3_4}) and eq.(\ref{eq:3_5}):
\begin{equation}\begin{aligned}
m^{2}_{H3-loop}=\frac{d^{2}V_{3-loop\;approx}}{d\varphi^{2}}|_{\varphi=v_{3}}=(\frac{I_{1}}{6}+\frac{I_{2}}{24}+\frac{I_{2}}{12}\ln\frac{v^{2}_{3}}{\mu^{2}}+\frac{I_{3}}{48}(\ln\frac{v^{2}_{3}}{\mu^{2}}+\ln\frac{v^{2}_{3}}{\mu^{2}})^{2})v^{2}_{3}.
\end{aligned}\label{eq:3_8}
\end{equation}\\
Equations (\ref{eq:3_6}), (\ref{eq:3_7}) and (\ref{eq:3_8}) give the Higgs boson mass in one-loop, two-loop and three-loop approximations, which are not the physical masses but the running masses.

\paragraph{ }In the massless case, we can simplify the condition in eq.(\ref{eq:3_1}) if we recall that $\mu$ is an arbitrary parameter, we are allowed to choose $\mu$ to be location of the minimum of the effective potential as has been done in ref.\cite{9p}. Equation (\ref{eq:3_2}), eq.(\ref{eq:3_3}), and eq.(\ref{eq:3_4}) become:

\begin{equation}
\frac{dV_{1-loop\;approx}}{d\varphi}|_{\varphi=v_{1}}=0\Rightarrow\lambda-\frac{11}{6}A_{1}=0,
\label{eq:3_9}
\end{equation}
\begin{equation}
\frac{dV_{2-loop\;approx}}{d\varphi}|_{\varphi=v_{2}}=0\Rightarrow\lambda-\frac{11}{6}(A_{1}+A)-\frac{35}{24}B=0,
\label{eq:3_10}
\end{equation}
\begin{equation}
\frac{dV_{3-loop\;approx}}{d\varphi}|_{\varphi=v_{3}}=0\Rightarrow\lambda-\frac{11}{6}I_{1}-\frac{35}{24}I_{2}-\frac{5}{12}I_{3}=0,
\label{eq:3_11}
\end{equation}\\
Note that, eq.(\ref{eq:3_9}) can be viewed as a generalization of the condition that relates the scalar coupling to the gauge coupling obtained in ref.\cite{9p} for the case of massless scalar quantum electrodynamics.\\\\
\newpage
The Higgs boson mass for $\mu=v$ is described by:

\begin{equation}
m^{2}_{H1-loop}=\frac{d^{2}V_{1-loop\;approx}}{d\varphi^{2}}|_{\varphi=v_{1}}=\frac{A_{1}}{6}v^{2}_{1}.
\label{eq:3_12}
\end{equation}

\begin{equation}
m^{2}_{H2-loop}=\frac{d^{2}V_{2-loop\;approx}}{d\varphi^{2}}|_{\varphi=v_{2}}=(\frac{A_{1}+A}{6}+\frac{B}{24})\;v^{2}_{2},
\label{eq:3_13}
\end{equation}

\begin{equation}
m^{2}_{H3-loop}=\frac{d^{2}V_{3-loop\;approx}}{d\varphi^{2}}|_{\varphi=v_{3}}=(\frac{A_{1}+A+K_{1}}{6}+\frac{B+K_{2}}{24})\;v^{2}_{2},
\label{eq:3_14}
\end{equation}

In present work, our initially input parameters are taken from ref.\cite{18p} at the neutral weak scale $M_{Z}$. These parameters can be computed at any scale for each loop by using the following perturbative expression:

\begin{equation}
g_{i}(t)=g_{i}(t=0)+\int (\beta^{(1)}_{g_{i}}(g_{i}(t))+\beta^{(2)}_{g_{i}}(g_{i}(t))+...)\;dt,
\label{eq:3_99}
\end{equation}
\begin{equation*}
(g_{i}=g_{1}, g_{2}, g_{3}, g_{t}),\;\; t=\ln \frac{\mu}{\mu_{0}},\; \text{with}\; \mu_{0}=M_{Z}\approx 91.19\;\text{GeV}.
\end{equation*}

\begin{figure}
\includegraphics{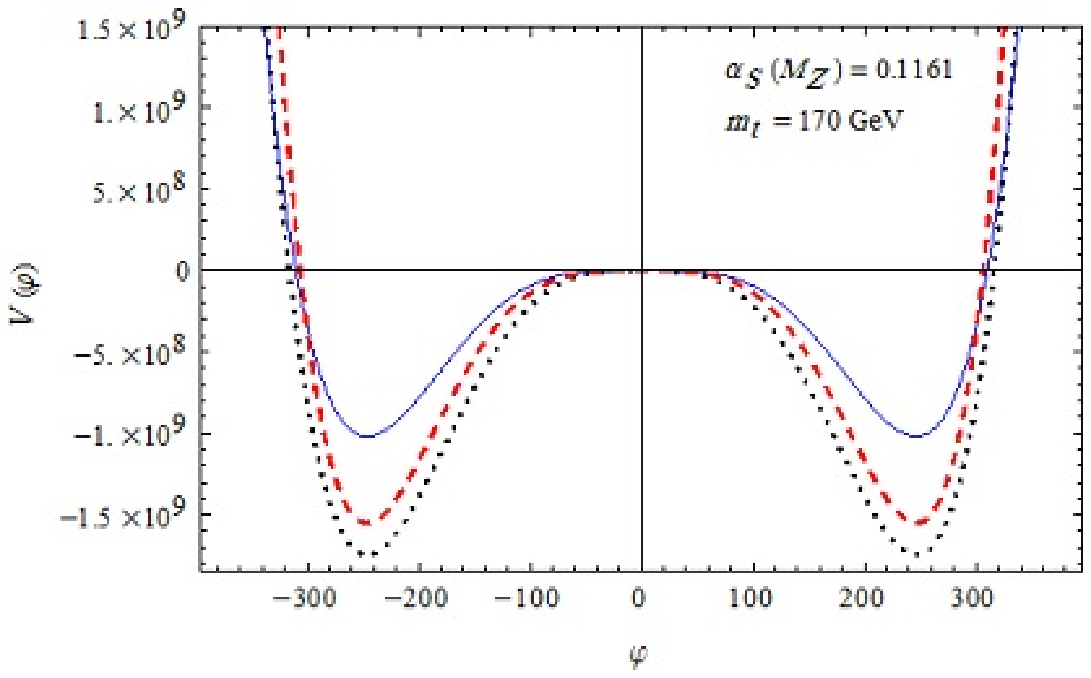}
\caption{A comparison between the one-loop (dotted line), two-loop (solid line) and three-loop (dashed line) effective potential for SM at $\mu$ scale =100 GeV.}
\label{fig:3_1}
\end{figure}

\begin{figure}
\includegraphics{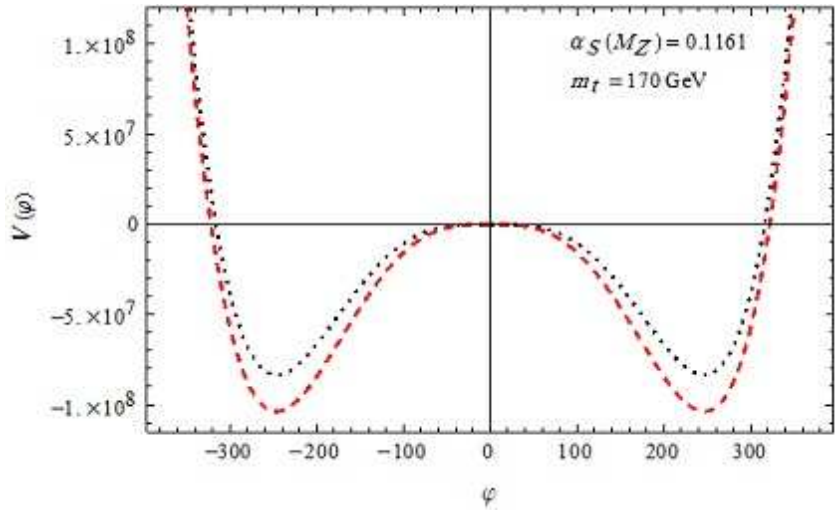}
\caption{A comparison between the one-loop (dotted line) and three-loop (dashed line) effective potential for SM at $\mu$ scale $=3\times 10^{3}$ GeV.}
\label{fig:3_3}
\end{figure}

For the strong coupling $\alpha_{S}$ and the top quark mass $m_{t}$ we have taken the range as in \cite{18p},\cite{14p}:
\begin{equation*}\begin{aligned}
0.1127\leq\alpha_{S}(M_{Z})\leq 0.1202,\\170\;\text{GeV}\leq m_{t}\leq 176\;\text{GeV}.
\end{aligned}\label{eq:3_122}
\end{equation*}\\
\textbf{\large Rusults and discussion}
\paragraph{ }The computations for the one-loop, two-loop and three-loop order within the $\mu-$range $(M_{Z}\leq \mu \leq 600 \text{GeV})$ are presented in Table.(\ref{tab:3_1}) for $m_{t}=170$ GeV, $\alpha_{S}(M_{Z})=0.1161$. In this table and the other tables we have used the reduced scalar coupling $\alpha_{\lambda}=\frac{\lambda}{4\pi}$, which corresponds to gauge couplings $\alpha_{g}=\frac{g^{2}}{4\pi}$, $g=g_{1}, g_{2}, g_{3}$ for U(1), SU(2) and SU(3), and appears as perturbative expansion parameter in the loop expansion of the effective potential. For the case $\mu=100$ GeV, there is a minimum of the one-loop, two-loop and three-loop effective potential (see Fig.(\ref{fig:3_1})) with too large values for Higgs scalar coupling and mass, so this result may be ruled out since it lies outside the range of validity of the approximation. It turns out that, increasing the value of $\mu$ improves the results. By considering the neutral weak mass $M_{Z}$ as a mass scale, the Higgs boson mass value is $m_{H1-loop}\approx 753$ GeV at one-loop order. This result is consistent with the recent LHC discovery \cite{19p}. However for the value, $\mu \approx 5.68\times 10^{2}$ GeV, we get $m_{H1-loop}\approx 239.2$ GeV, $m_{H2-loop}\approx 125.4$ GeV and $m_{H3-loop}\approx 181.8$ GeV. At  $\mu\geq 28\times 10^{2}\approx 11.4 \;\nu$ GeV, the scalar coupling $\lambda$ at two-loop becomes negative and leads to metastable vacuum while the three-loop order is stable even at high scale energy $\approx 10^{19}$ GeV. For the larger $\mu$-range $(3 \times 10^{3} \text{GeV}\leq \mu \leq 20 \times 10^{3} \text{GeV})$ at $m_{t}=170$ GeV, $\alpha_{S}(M_{Z})=0.1161$ spontaneous symmetry breaking for one-loop and three-loop occurs (see Fig.(\ref{fig:3_3}) at approximately the same scalar coupling values as shown in Table.(\ref{tab:3_2}). In addition, we get $m_{H1-loop}\approx 126.14$ GeV at $\mu \approx 6.5 \times 10^{3}$ GeV while for three-loop $m_{H3-loop}\approx 126$ GeV at $\mu \approx 20 \times 10^{3}$ GeV.

\paragraph{ }We present our results in Table.(\ref{tab:3_3}) for the scalar Higgs coupling, obtained from eqs.(\ref{eq:3_9}),(\ref{eq:3_10}) and eq.(\ref{eq:3_11}), and Higgs boson mass in the ranges of $0.1127\leq\alpha_{S}\leq 0.1202$ and $170\;\text{GeV}\leq m_{t}\leq 176\; \text{GeV}$ for each loop order with consideration that the mass scale equals to vacuum expectation value $(\mu=v)$. Thus the Higgs boson mass for each loop is:
\begin{equation}\begin{aligned}
346.6\; \text{GeV}\leq m_{H1-loop}\leq 347.2 \;\text{GeV},\\
241.2\; \text{GeV}\leq m_{H2-loop}\leq 241.5 \;\text{GeV},\\
247.6\; \text{GeV}\leq m_{H3-loop}\leq 247.9 \;\text{GeV}.
\end{aligned}\label{eq:3_144}
\end{equation}\\
At first glance to the numerical results of one-loop order, one can find that Higgs boson mass $m_{H}$  is large for all the range under our study because the values of the scalar Higgs coupling  $\lambda$ may be too large to justify the neglect of higher order contributions. However, the two-loop and three-loop results are not appreciably different from each other. The two-loop and three-loop radiative corrections have led to an improvement of the Higgs boson mass and the value of the scalar coupling.
\paragraph{ }It is interesting to note that in massless theory the absence of tree level mass term leads to large value for the scalar coupling to the limit, where the perturbation method becomes invalid, since as known the perturbation parameters should be small in order to become successively smaller at higher-order terms. In order to overcome this problem one can choose a suitable mass scale with the inclusion of higher-order contributions.

\paragraph{ }The recent discovery of Higgs-like particle does not confirm or exclude the possibility of many Higgs particles with masses larger than the one discovered in CERN. Extensions of the standard model to for example supersymmetry (SUSY) models predict the existence of families of Higgs bosons, rather than the one Higgs particle of the standard model.  In all of them, there is one neutral Higgs boson with properties similar to those of the standard model Higgs boson in addition to neutral and charged Higgs bosons. For example in the Minimal Supersymmetric Standard Model (MSSM), which the simplest model of the SUSY, the Higgs mechanism yields to two Higgs doublets, leading to the existence of a quintet of scalar particles. In this model, the lightest Higgs boson mass that represent the standard model-like Higgs, is predicted to be $m_{H}<135$ GeV. Conversely, larger values of the standard model-like Higgs mass up to $\approx 250$ GeV can be obtained with the Non-Minimal Supersymmetric Standard Model (NMSSM) that can avoid the problems of the MSSM with larger particle content \cite{22p},\cite{20p}. 

\paragraph{ }Recently, late last year (2015), a heavy Higgs-like particle is not exclude from the Large Hadron Collider LHC where ATLAS and CMS at LHC have reported that proton-to-proton collisions had led to create more photon pairs with energies around 750 GeV than was expected. This excess in the diphoton mass distribution ($m_{\gamma\gamma}\approx 750$ GeV), if confirmed, would indicate the presence of unexpected new elementary particles \cite{19p}. A large number of theoretical interpretations have appeared following LHC discovery \cite{21p,29p,19p}. Most of these papers explain the excess through some new boson with mass-750 GeV. For example in ref.\cite{29p}, a heavy Higgs-like boson which is six times heavier than the standard model-Higgs particle is expected to couple with new types of fermions.

\paragraph{ }In addition, there is an agreement between our result eq.(\ref{eq:3_144}) and some pervious theoretical results of other authors employing different methods \cite{23p,24p,25p,26p,27p,28p}.

\begin{figure}
\includegraphics{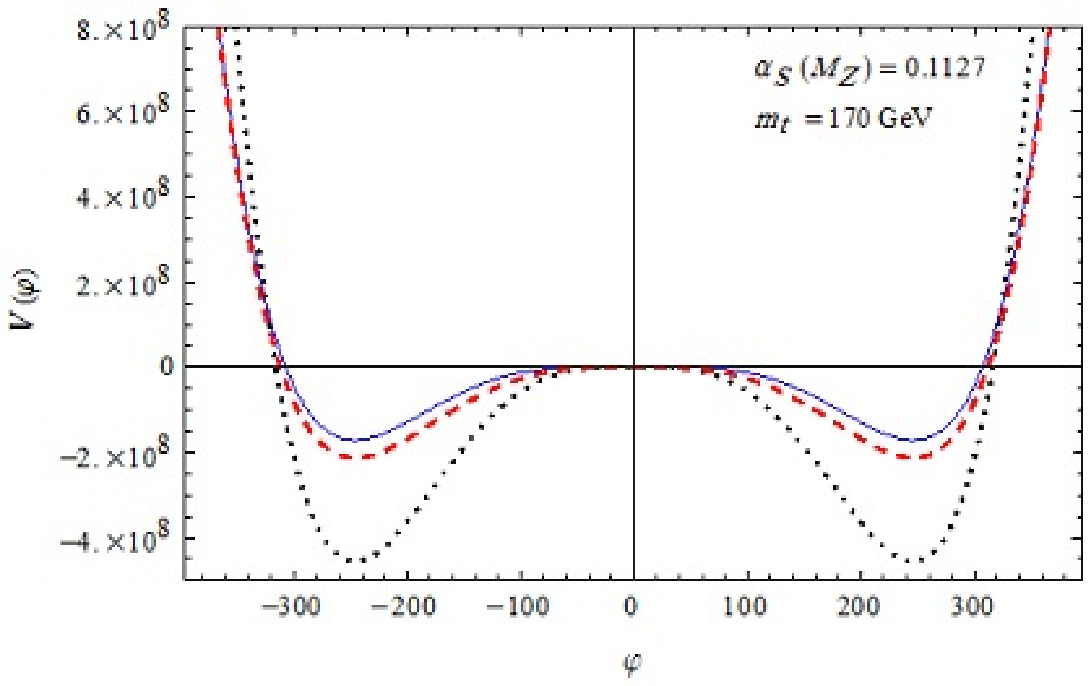}
\caption{A comparison between the one-loop (dotted line), two-loop (solid line) and three-loop(dashed line) effective potential for SM at $\mu$ scale =$v$.}
\label{fig:3_5}
\end{figure}

\begin{figure}
\includegraphics{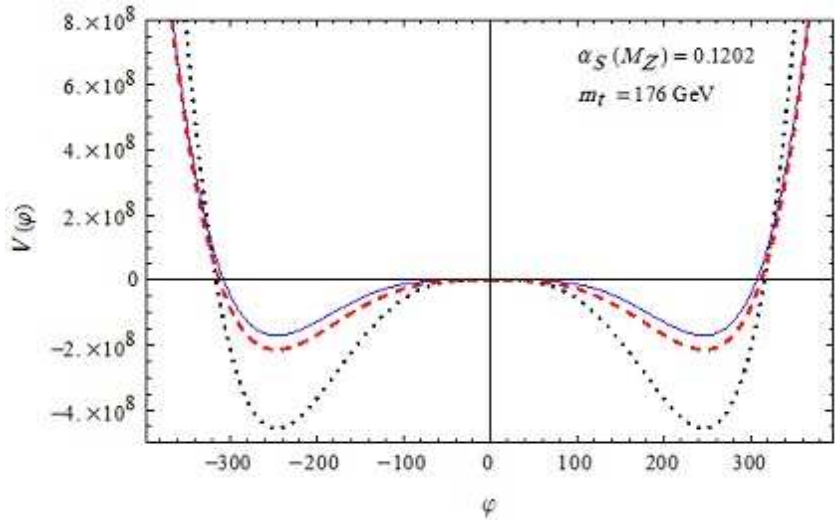}
\caption{A comparison between the one-loop (dotted line), two-loop (solid line) and three-loop(dashed line) effective potential for SM at $\mu$ scale =$v$.}
\label{fig:3_8}
\end{figure}

\paragraph{ }A comparison between the one, two and three-loop effective potential at lowest values $m_{t}=170$ GeV and $\alpha_{S}(M_Z )=0.1127$ and largest values $m_{t}=176$ GeV and $\alpha_{S}(M_Z )=0.1202$ are shown in Figs.(\ref{fig:3_5}) and (\ref{fig:3_8}). The conclusion that can be inferred from these figures is the curves of the two-loop and three-loop effective potential are almost similar and the position of the vacuum for the three-loop order is slightly deeper than the two-loop. The one-loop effective potential is deeper at the vacuum position although it has the same trend. Furthermore, it must be noted that all these figures with different values range of $m_{t}$ and $\alpha_{S}$ are almost identical and show similar trend.

\newpage

\begin{longtable}[c]{| c | c | c | c | c | c | c | c | c | c |}

\caption{Values of the Higgs scalar coupling, $\alpha_{\lambda}=\lambda/4\pi$, and Higgs boson mass for one-loop, two-loop and three-loop order at $m_{t}=170$ GeV and $\alpha_{S}(M_{Z})=0.1161$ for the mass scale-range $100\; \text{GeV}\leq\mu\leq600\; \text{GeV}$.\label{tab:3_1}}\\
 \hline
\multicolumn{1}{| c |}{$\mu\text{(GeV)}$}&\multicolumn{2}{| c |}{one-loop}&\multicolumn{2}{|c|}{two-loop}&\multicolumn{2}{| c |}{three-loop}\\ 
{}&$\alpha_{\lambda}$&{$m_{H}$(GeV)}&{$\alpha_{\lambda}$}&{$m_{H}$(GeV)}&{$\alpha_{\lambda}$}&{$m_{H}$(GeV)}\\ \hline
$M_{Z}$&3.75&753.1&2.14&650.71&2.67&1053.13\\
100&3.39&678.74&1.95&563.87&2.17&731.48\\
150&2.37&473.87&1.50&357.90&1.54&371.45\\
200&1.95&390.59&1.34&281.76&1.36&285.41\\
250&1.72&343.88&1.25&238.77&1.26&245.48\\
300&1.57&313.37&1.19&209.60&1.20&222.62\\
350&1.46&291.53&1.15&187.69&1.16&208.26\\
400&1.38&274.83&1.13&169.95&1.12&198.64\\
450&1.32&261.82&1.11&154.87&1.10&191.82\\
500&1.26&251.06&1.09&141.51&1.07&186.60\\
550&1.22&242.07&1.08&129.42&1.05&182.90\\
600&1.18&234.53&1.07&117.87&1.03&179.82\\
\hline
\end{longtable}

\begin{longtable}[c]{| c | c | c | c | c | c | c | c | c | c |}

\caption{Values of the Higgs scalar coupling, $\alpha_{\lambda}=\lambda/4\pi$, and Higgs boson mass for one-loop and three-loop order at $m_{t}=170$ GeV and $\alpha_{S}(M_{Z})=0.1161$ for the large mass scale-range $3\times 10^{3}\; \text{GeV}\leq\mu\leq 20\times 10^{3}\; \text{GeV}$.\label{tab:3_2}}\\
 \hline
\multicolumn{1}{| c |}{$\mu\text{(GeV)}$}&\multicolumn{2}{| c |}{one-loop}&\multicolumn{2}{| c |}{three-loop}\\ 
{}&$\alpha_{\lambda}$&{$m_{H}$(GeV)}&{$\alpha_{\lambda}$}&{$m_{H}$(GeV)}\\ \hline

$3\times 10^{3}$&0.75&148.42&0.75&154.49\\
$6\times 10^{3}$&0.65&128.11&0.66&144.03\\
$9\times 10^{3}$&0.60&118.64&0.62&138.9\\
$12\times 10^{3}$&0.57&112.67&0.59&134.28\\
$15\times 10^{3}$&0.55&108.43&0.57&131.7\\
$18\times 10^{3}$&0.54&105.31&0.55&129.1\\
$20\times 10^{3}$&0.53&103.43&0.54&126\\

\hline
\end{longtable}

\begin{longtable}[c]{| c | c | c | c | c | c | c | c | c | c | c | c |}

\caption{Values of the Higgs scalar coupling, $\alpha_{\lambda}=\lambda/4\pi$, and Higgs boson mass for one-loop, two-loop and three-loop order with different values of the top quark mass $m_{t}$ and the strong coupling $\alpha_{S}(M_{Z})$ in the mass scale $\mu=v$.\label{tab:3_3}}\\
 \hline
\multicolumn{1}{| c |}{$\alpha_{S}(M_{Z})$}&\multicolumn{1}{| c |}{$m_{t}$(GeV)}&\multicolumn{2}{c}{one-loop}&\multicolumn{2}{| c |}{two-loop}&\multicolumn{2}{| c |}{three-loop}\\ 
{}&{}&{$\alpha_{\lambda}$}&{$m_{H}$(GeV)}&{$\alpha_{\lambda}$}&{$m_{H}$(GeV)}&{$\alpha_{\lambda}$}&{$m_{H}$(GeV)}\\ \hline

0.1127&170&1.737&346.75&1.252&241.38&1.269&247.66\\
{}&172&1.739&346.94&1.254&241.40&1.271&247.84\\
{}&174&1.740&347.00&1.255&241.40&1.273&247.84\\
{}&176&1.741&347.16&1.256&241.24&1.275&247.84\\
\hline

0.1161&170&1.738&346.77&1.256&241.44&1.269&247.72\\
{}&172&1.738&346.81&1.254&241.46&1.271&247.74\\
{}&174&1.740&347.00&1.255&241.34&1.272&247.75\\
{}&176&1.741&347.19&1.256&241.34&1.275&247.92\\
\hline

0.118&170&1.738&346.78&1.253&241.36&1.268&247.59\\
{}&172&1.738&346.82&1.253&241.38&1.271&247.78\\
{}&174&1.740&347.01&1.255&241.34&1.272&247.79\\
{}&176&1.740&347.04&1.256&241.34&1.274&247.80\\
\hline

0.1202&170&1.736&346.63&1.252&241.51&1.268&247.64\\
{}&172&1.738&346.83&1.254&241.37&1.270&247.65\\
{}&174&1.740&347.03&1.254&241.38&1.271&247.66\\
{}&176&1.741&347.06&1.256&241.23&1.274&247.83\\
\hline
\end{longtable}

\paragraph{}

\newpage
\section{Conclusion}
\paragraph{ } In this paper, we have used the renormalization group method to calculate the effective potential for massless standard model up to three-loop order. We have also investigated the stability of the effective potential. It is shown that while the vacuum structure for two-loop order is metastable, it is stable for three-loop order even at high energy scale. The Higgs boson mass is calculated at each loop order for a range of energy scale and a range of the top quark mass and the strong coupling. The obtained values of the Higgs boson mass are in the experimental range for the scale energy of order $20\times 10^{3}$ GeV.

\newpage
\begin{appendices}
\renewcommand{\theequation}{\arabic{equation}}

\section{}
\begin{equation}\begin{aligned}
\gamma^{(1)}=\frac{1}{4k}(3 g^{2}_{1}+9 g^{2}_{2}-12 g^{2}_{t}),
\end{aligned}
\end{equation}

\begin{equation}\begin{aligned}
\beta^{(1)}_{\lambda}=\frac{1}{k}(4\lambda^{2}-\lambda(3 g^{2}_{1}+9 g^{2}_{2}-12 g^{2}_{t})+\frac{9}{4}g^{4}_{1}+\frac{9}{2}g^{2}_{1}g^{2}_{2}+\frac{27}{4}g^{4}_{2}-36 g^{4}_{t}),
\end{aligned}
\end{equation}

\begin{equation}\begin{aligned}
\beta^{(1)}_{g_{1}}=\frac{41}{6k}g^{3}_{1},\;\;\beta^{(1)}_{g_{2}}=\frac{-19}{6k}g^{3}_{2},\;\; \beta^{(1)}_{g_{3}}=\frac{-7}{k}g^{3}_{3},
\end{aligned}
\end{equation}

\begin{equation}\begin{aligned}
\beta^{(1)}_{g_{t}}=\frac{1}{k}(\frac{9}{2}g^{3}_{t}-8 g^{2}_{3}g_{t}-\frac{17}{2}g_{t}g^{2}_{1}),
\end{aligned}
\end{equation}

\begin{equation}\begin{aligned}
\gamma^{(2)}=\frac{-1}{6 k^{2}}\lambda^{2}-\frac{1}{k^{2}}(\frac{431}{96}g^{4}_{1}+\frac{9}{2}g^{2}_{1}g^{2}_{2}-\frac{271}{32}g^{4}_{2}+\frac{85}{24}g^{2}_{1}g^{2}_{t}+\frac{45}{8}g^{2}_{2}g^{2}_{t}+20 g^{2}_{3}g^{2}_{t}-\frac{27}{4}g^{4}_{t}),
\end{aligned}
\end{equation}

\begin{equation}\begin{aligned}
\beta^{(2)}_{\lambda}=\frac{1}{ k^{2}}[\frac{-26}{3}\lambda^{3}+\lambda^{2}(6 g^{2}_{1}+18 g^{2}_{2}-24 g^{2}_{t})\;\;\;\;\;\;\;\;\;\;\;\;\;\;\;\;\;\;\;\;\;\;\;\;\;\;\;\;\;\;\;\;\;\;\;\;\;\;\;\;\;\;\;\;\;\;\;\;\;\;\;\;\;\;\;\;\;\;\;\;\;\;\;\;\;\;\;\\+\lambda(\frac{629}{24}g^{4}_{1}+\frac{39}{4}g^{2}_{1}g^{2}_{2}+\frac{85}{6}g^{2}_{1}g^{2}_{t}-\frac{73}{8}g^{4}_{2}+\frac{45}{2}g^{2}_{2}g^{2}_{t}+80 g^{2}_{3}g^{2}_{t}-3 g^{4}_{t})+180 g^{6}_{t}-192 g^{2}_{3}g^{4}_{t}\\-16 g^{2}_{1}g^{4}_{t}-\frac{27}{2}g^{4}_{2}g^{2}_{t}+63\;g^{2}_{1}g^{2}_{t}g^{2}_{2}-\frac{57}{2}g^{4}_{1}g^{2}_{t}-\frac{379}{8}g^{6}_{1}-\frac{559}{8}g^{4}_{1}g^{2}_{2}-\frac{289}{8}g^{2}_{1}g^{4}_{2}+\frac{915}{8}g^{6}_{2}],
\end{aligned}
\end{equation}

\begin{equation}\begin{aligned}
\beta^{(2)}_{g_{1}}=\frac{1}{k^{2}} g^{3}_{1}[\frac{199}{18}g^{2}_{1}+\frac{9}{2}g^{2}_{2}+\frac{44}{3}g^{2}_{3}-\frac{17}{6}g^{2}_{t}],
\end{aligned}
\end{equation}
\begin{equation}\begin{aligned}
\beta^{(2)}_{g_{2}}=\frac{1}{k^{2}} g^{3}_{2}[\frac{3}{2}g^{2}_{1}+\frac{35}{6}g^{2}_{2}+12 g^{2}_{3}-\frac{3}{2}g^{2}_{t}],
\end{aligned}
\end{equation}
\begin{equation}\begin{aligned}
\beta^{(2)}_{g_{3}}=\frac{1}{k^{2}} g^{3}_{3}[\frac{11}{6}g^{2}_{1}+\frac{9}{2}g^{2}_{2}-26 g^{2}_{3}-2 g^{2}_{t}],
\end{aligned}
\end{equation}
\begin{equation}\begin{aligned}
\beta^{(2)}_{g_{t}}=\frac{1}{k^{2}}g_{t}[-12 g^{4}_{t}+g^{2}_{t}(\frac{131}{16}g^{2}_{1}+\frac{225}{16}g^{2}_{2}+36 g^{2}_{3}-2\lambda)+\frac{1187}{216}g^{4}_{1}-\frac{3}{4}g^{2}_{1}g^{2}_{2}+\frac{19}{9}g^{2}_{3}g^{2}_{1}\\-\frac{23}{4}g^{4}_{2}+9 g^{2}_{3}g^{2}_{2}-108 g^{4}_{3}+\frac{1}{6}\lambda^{2}],\;\;\;\;\;\;\;\;\;\;\;\;\;\;\;\;\;\;\;\;\;\;\;\;\;\;\;\;\;\;\;\;\;\;\;\;\;\;\;\;\;\;\;\;\;\;\;\;\;\;\;\;\;\;\;\;
\end{aligned}
\end{equation}
\\
\begin{equation}\begin{aligned}
\gamma^{(3)}=\frac{-1}{k^{3}}[\frac{125}{27}g^{6}_{1}(\frac{27053}{2000}-\frac{13791}{2000}\zeta(3))+\frac{25}{3}g^{4}_{1}g^{2}_{3}(\frac{99}{20}-\frac{132}{25}\zeta(3))+g^{6}_{2}(\frac{6785}{576}+\frac{1953}{16}\zeta(3))\\+\frac{25}{9}g^{4}_{1}g^{2}_{2}(\frac{1377}{800}-\frac{1215}{400}\zeta(3))+\frac{5}{3}g^{2}_{t}g^{2}_{1}g^{2}_{2}(\frac{1113}{320}+\frac{81}{10}\zeta(3))+\frac{5}{3}g^{2}_{1}g^{4}_{2}(\frac{1629}{320}-\frac{459}{80}\zeta(3))\\+\frac{5}{3}g^{2}_{t}g^{2}_{1}g^{2}_{3}(\frac{-2419}{60}+\frac{204}{5}\zeta(3))+\frac{25}{9}g^{2}_{t}g^{4}_{1}(\frac{-144271}{9600}+\frac{3}{100}\zeta(3))+g^{2}_{t}g^{4}_{2}(\frac{4275}{128}-\frac{189}{4}\zeta(3))\\+\frac{5}{3}g^{2}_{1}g^{2}_{2}\frac{\lambda}{6}(\frac{117}{40}-\frac{27}{5}\zeta(3))+\frac{25}{9}g^{4}_{1}\frac{\lambda}{6}(\frac{351}{400}-\frac{81}{50}\zeta(3))+ g^{4}_{2}\frac{\lambda}{6}(\frac{117}{16}-\frac{27}{2}\zeta(3))\;\;\;\;\;\;\;\;\;\;\;\;\;\;\;\;\;\\+\frac{5}{3}g^{4}_{t}g^{2}_{1}(\frac{-957}{80}-\frac{9}{5}\zeta(3))+g^{2}_{t}g^{4}_{3}(\frac{622}{3}-24\zeta(3))+g^{2}_{t}g^{2}_{2}g^{2}_{3}(\frac{-489}{4}+108\zeta(3))\;\;\;\;\;\;\;\;\;\;\;\;\;\;\;\\+3g^{4}_{2}g^{2}_{3}(\frac{135}{4}-36\zeta(3))+g^{4}_{t}g^{2}_{2}(\frac{-1161}{16}+27\zeta(3))+g^{4}_{t}g^{2}_{3}(\frac{15}{2}-27\zeta(3))+g^{6}_{t}(\frac{789}{16}+9\zeta(3))\\+45\frac{\lambda}{6} g^{4}_{t}+45\frac{\lambda^{2}}{36}g^{2}_{2}+15\frac{\lambda^{2}}{36}g^{2}_{1}-36\frac{\lambda^{3}}{216}-\frac{135}{2}\frac{\lambda^{2}}{36}g^{2}_{t}],\;\;\;\;\;\;\;\;\;\;\;\;\;\;\;\;\;\;\;\;\;\;\;\;\;\;\;\;\;\;\;\;\;\;\;\;\;\;\;\;\;\;\;\;\;\;\;\;\;\;\;\;\;\;\;\;\;\;
\end{aligned}
\end{equation}

\begin{equation}\begin{aligned}
\beta^{(3)}_{\lambda}=\frac{12}{k^{3}}[g^{8}_{2}(\frac{228259}{3072}-\frac{20061}{128}\zeta(3))+g^{2}_{1}g^{6}_{2}(\frac{-165665}{3456}-\frac{405}{32}\zeta(3))+g^{4}_{1}g^{4}_{2}(\frac{-81509}{3456}+\frac{2217}{64}\zeta(3))\;\;\;\;\;\;\;\;\;\;\;\;\;\;\;\;\;\;\;\;\;\;\;\;\;\;\;\;\;\;\;\\+g^{6}_{1}g^{2}_{2}(\frac{-237787}{6912}+\frac{2177}{96}\zeta(3))+g^{8}_{1}(\frac{-466605}{9216}+\frac{12457}{384}\zeta(3))+\frac{1}{6}\lambda g^{6}_{2}(\frac{58031}{288}+\frac{4419}{8}\zeta(3))\;\;\;\;\;\;\;\;\;\;\;\;\;\;\;\;\;\;\;\;\;\;\;\;\;\;\;\;\;\;\;\;\\+\frac{1}{6}\lambda g^{2}_{1}g^{4}_{2}(\frac{6137}{32}-\frac{393}{8}\zeta(3))+\frac{1}{6}\lambda g^{4}_{1}g^{2}_{2}(\frac{1549}{8}-\frac{147}{8}\zeta(3))+\frac{1}{6}\lambda g^{6}_{1}(\frac{88639}{432}-\frac{1493}{24}\zeta(3))\;\;\;\;\;\;\;\;\;\;\;\;\;\;\;\;\;\;\;\;\;\;\;\;\;\;\;\;\;\;\;\\+\frac{1}{36}\lambda^{2}g^{4}_{2}(\frac{-1389}{8}-513\zeta(3))+\frac{1}{36}\lambda^{2}g^{2}_{1}g^{2}_{2}(-333-162\zeta(3))+\frac{1}{36}\lambda^{2}g^{4}_{1}(-418-81\zeta(3))\;\;\;\;\;\;\;\;\;\;\;\;\;\;\;\;\;\;\;\;\;\;\;\;\;\;\;\;\;\;\;\\+\frac{1}{216}\lambda^{3}g^{2}_{2}(-474+72\zeta(3))+\frac{1}{216}\lambda^{3}g^{2}_{1}(-158+24\zeta(3))+\frac{1}{1296}\lambda^{4}(3588+2016\zeta(3))\;\;\;\;\;\;\;\;\;\;\;\;\;\;\;\;\;\;\;\;\;\;\;\;\;\;\;\;\;\;\;\\+g^{2}_{t}g^{6}_{1}(\frac{125503}{2304}-\frac{5}{2}\zeta(3))+g^{2}_{t}g^{6}_{2}(\frac{-6849}{256}+\frac{297}{4}\zeta(3))+g^{2}_{t}g^{2}_{1}g^{4}_{2}(\frac{3487}{256}+\frac{27}{4}\zeta(3))\;\;\;\;\;\;\;\;\;\;\;\;\;\;\;\;\;\;\;\;\;\;\;\;\;\;\;\;\;\\+g^{2}_{t}g^{4}_{1}g^{2}_{2}(\frac{25441}{768}-3\zeta(3))+\frac{1}{6}\lambda g^{2}_{t}g^{4}_{2}(\frac{-6957}{64}-\frac{351}{2}\zeta(3))+\frac{1}{6}\lambda g^{2}_{t}g^{2}_{1}g^{2}_{2}(\frac{-6509}{32}+177\zeta(3))\;\;\;\;\;\;\;\;\;\;\;\;\;\;\;\;\;\;\;\;\;\;\;\;\;\;\;\;\;\\+\frac{1}{6}\lambda g^{2}_{t}g^{4}_{1}(\frac{-203887}{1728}-\frac{449}{6}\zeta(3))+\frac{1}{36}\lambda^{2}g^{2}_{t}g^{2}_{2}(\frac{639}{4}-432\zeta(3))+\frac{1}{36}\lambda^{2}g^{2}_{t}g^{2}_{1}(\frac{-195}{4}-48\zeta(3))\;\;\;\;\;\;\;\;\;\;\;\;\;\;\;\;\;\;\;\;\;\;\;\;\;\;\\+\frac{97}{24}\lambda^{3}g^{2}_{t}+g^{4}_{t}g^{4}_{2}(\frac{9909}{128}-\frac{819}{16}\zeta(3))+g^{4}_{t}g^{2}_{1}g^{2}_{2}(\frac{-1079}{192}-\frac{743}{8}\zeta(3))+g^{4}_{t}g^{4}_{1}(\frac{67793}{3456}+\frac{2957}{144}\zeta(3))\;\;\;\;\;\;\;\;\;\;\;\;\;\;\;\;\;\;\;\;\;\;\;\;\\+\frac{1}{6}\lambda g^{4}_{t}g^{2}_{2}(\frac{-4977}{8}+513\zeta(3))+\frac{1}{6}\lambda g^{4}_{t}g^{2}_{1}(\frac{-2485}{24}+57\zeta(3))+\frac{1}{36}\lambda^{2}g^{4}_{t}(\frac{1719}{2}+756\zeta(3))\;\;\;\;\;\;\;\;\;\;\;\;\;\;\;\;\;\;\;\;\;\;\;\;\;\\+g^{6}_{t}g^{2}_{2}(\frac{3411}{32}-27\zeta(3))+g^{6}_{t}g^{2}_{1}(\frac{3467}{96}+17\zeta(3))+\frac{1}{6}\lambda g^{6}_{t}(\frac{117}{8}-198\zeta(3))+g^{8}_{t}(\frac{-1599}{8}-36\zeta(3))\;\;\;\;\;\;\;\;\;\;\;\;\;\;\;\;\;\;\;\;\;\;\;\\+g^{2}_{3}g^{6}_{2}(\frac{-459}{8}+54\zeta(3))+g^{2}_{3}g^{2}_{1}g^{2}_{4}(\frac{-153}{8}+18\zeta(3))+g^{2}_{3}g^{4}_{1}g^{2}_{2}(\frac{-187}{8}+22\zeta(3))+g^{2}_{3}g^{6}_{1}(\frac{-187}{8}+22\zeta(3))\;\;\;\;\;\;\;\;\;\;\;\;\;\;\;\;\;\;\;\;\;\;\;\;\\+\frac{1}{6}\lambda g^{2}_{3}g^{4}_{2}(\frac{405}{2}-216\zeta(3))+\frac{1}{6}\lambda g^{2}_{3}g^{4}_{1}(\frac{165}{2}-88\zeta(3))+g^{2}_{3}g^{2}_{t}g^{4}_{2}(\frac{651}{8}-54\zeta(3))+g^{2}_{3}g^{2}_{1}g^{2}_{t}g^{2}_{2}(\frac{249}{4}-36\zeta(3))\;\;\;\;\;\;\;\;\;\;\;\;\;\;\;\;\;\;\;\;\;\;\;\\+g^{2}_{3}g^{4}_{1}g^{2}_{t}(\frac{587}{24}-18\zeta(3))+\frac{1}{6}\lambda g^{2}_{3}g^{2}_{2}g^{2}_{t}(\frac{-489}{2}+216\zeta(3))+\frac{1}{6}\lambda g^{2}_{3}g^{2}_{1}g^{2}_{t}(\frac{-2419}{18}+136\zeta(3))\;\;\;\;\;\;\;\;\;\;\;\;\;\;\;\;\;\;\;\;\;\;\;\;\;\;\;\;\;\;\;\;\;\\+\frac{1}{36}\lambda^{2}g^{2}_{t}g^{2}_{3}(-1224+1152\zeta(3))+g^{2}_{3}g^{2}_{1}g^{4}_{t}(\frac{931}{18}-\frac{56}{3}\zeta(3))+g^{2}_{3}g^{2}_{2}g^{4}_{t}(\frac{-31}{2}+24\zeta(3))+\frac{1}{6}\lambda g^{2}_{3}g^{4}_{t}(895-1296\zeta(3))\;\;\;\;\;\;\;\;\;\;\;\;\;\;\\+g^{2}_{3}g^{6}_{t}(-38+240\zeta(3))+\frac{1}{6}\lambda g^{4}_{3}g^{2}_{t}(\frac{1244}{3}-48\zeta(3))+g^{4}_{t}g^{4}_{3}(\frac{-266}{3}+32\zeta(3))],\;\; k=16 \pi^{2}.\;\;\;\;\;\;\;\;\;\;\;\;\;\;\;\;\;\;\;\;\;\;\;\;\;\;\;\;\;\;\;\;\;\;\;\;\;\;\;\;
\end{aligned}
\end{equation}
\end{appendices}

\begin{thebibliography}{99}
\bibitem {1p} G. 'tHooft, Nucl. Phys. \underline{B35} (1971) 167.
\bibitem {2p}J. Goldstone, A. Salam and S. Weinberg, Phys. Rev. \underline{127} (1962) 965.
\bibitem {3p}G. Jona-Lasinio, Nuovo Cimento \underline{34} (1964) 1790.
\bibitem {4p}B. Zumino, (1970) in Lectures in\textit{Elementary Particles and Quantum Field Theory}, edit by S. Deser et al, MIT Cambridge, Mass.
\bibitem {5p}K. Symanzik, Commun. Math. Phys. \underline{16} (1970) 48.
\bibitem {6p}R. H. Brandenberger, Rev. Mod. Phys. \underline{57} (1985) 1.
\bibitem {7p}M. Sher, Phys. Rep. \underline{179} (1989) 274.
\bibitem {9p}S. Coleman and E. Weinberg, Phys. Rev. \underline{D7} (1973) 1888.
\bibitem {8p}G.'tHooft, Nucl. Phys. \underline{B61} (1973) 455; S. Weinberg, Phys. Rev. \underline{D8} (1973) 3497.
\bibitem {10p}H. Alhendi, Phys. Rev. \underline{D37} (1988) 3749; ibdi \underline{40} (1989) 683 E.
\bibitem {11p}M. Machacek and M. T. Vaughn, Nucl. Phys. \underline{B249} (1985) 70.
\bibitem {12p}C. Ford, I. Jack and D. R. T. Jones, Nucl. Phys. \underline{B387} (1992) 373; \underline{B504} (1997) 551(E).
\bibitem {13p}M. Luo and Y. Xiao, Phys. Rev. Lett. \underline{90} (2003) 011601.
\bibitem {14p}H. A. Alhendi, T. Barakat and I. Gh. Loqman, Phys. Rev. \underline{D82} (2010) 053008.
\bibitem {15p}A. V. Bednyakov, A. F. Pikelner and V. N. Velizhanin, Nucl. Phys. \underline{B875} (2013) 552; Phys. Lett. \underline{B722} (2013) 336.
\bibitem {16p}K. G. Chetyrkin and M. F. Zoller, JHEP. \underline{1304} (2013) 091.
\bibitem {17p}I. Masina, Phys. Rev. \underline{D87} (2013) 053001.
\bibitem {18p}C. Amsler et al, "\textit{Particle Data Group Collaboration}", Phys. Lett. \underline{B667} (2008) 1.
\bibitem {19p}R. Garisto, Phys. Rev. Lett. \underline{116} (2016) 150001.
\bibitem {22p}K.A. Olive et al, (Particle Data Group), Chin. Phys. \underline{C38} (2014) 090001, and 2015 update.
\bibitem {20p}N. Christensen, T. Han and S. Su, Phys. Rev. \underline{D85} (2012) 115018; U. Ellwanger, C. Hugonie and A. M. Teixeira, Phys. Rept.  \underline{496} (2010) 1.
\bibitem {21p}Y. Nakiai, R. Sato and K. Tobioka, Phys. Rev. Lett. \underline{116} (2016) 151802; C. Petersson and R. Torre, Phys. Rev. Lett. \underline{116} (2016) 151804; W. Cho et al, Phys. Rev. Lett.\underline{116} (2016) 151805.
\bibitem {29p}G. Li et al, Phys. Rev. Lett. \underline{116} (2016) 151803.
\bibitem {23p}V. Elias, R. B. Mann, D. G. C. McKeon and T. G. Steele, Phys. Rev. Lett. \underline{91} (2003) 251601.
\bibitem {24p}F. A. Chishtie, V. Elias and T. G. Steele, Int. J. Mod. Phys. \underline{A20} (2005) 6241.
\bibitem {25p}P. P. Pal and S. Chakrabarty, Acta. Phys. Pol. \underline{B40} (2009) 1645.
\bibitem {26p}P. Kielanowski, S. R. Juarez W and H. G. Solis-Rodriguez, Phys. Rev. \underline{D72} (2005) 096003.
\bibitem {27p}M. D. Scadron, R. Delbourgo and G. Rupp, J. Phys. \underline{G32} (2006) 735.
\bibitem {28p}V. A. Bednyakov, N. D. Giokaris and A. V. Bednyakov, Phys. Part. Nucl. \underline{39} (2008) 13.
\end{thebibliography}
\end{document}